\documentclass[twocolumn,prb,aps,epsf,showpacs,superscriptaddress]{revtex4-1}

\usepackage{graphicx}
\usepackage{dcolumn}
\usepackage{bm}

\newcommand{\I}{\mathrm{i}}

\newcommand{\dxxyy}{$d_{x^2 - y^2}$}
\newcommand{\dwave}{$d$-wave}
\newcommand{\swave}{$s$-wave}

\newcommand{\hata}{$\hat a$}
\newcommand{\hatb}{$\hat b$}
\newcommand{\tc}{$T_\mathrm{c}$}

\newcommand{\lamb}{$\lambda$}
\newcommand{\cecoin}{CeCoIn$_5$}

\newcommand{\ybco}[1]{YBa$_2$Cu$_3$O$_{#1}$}

\newcommand{\bscco}{Bi$_2$Sr$_2$CaCu$_2$O$_{6 + \delta}$}
\newcommand{\tbcod}{Tl$_2$Ba$_2$CuO$_{6+\delta}$}

\newcommand{\Br}{\mbox{$\kappa$-(BEDT-TTF)$_2$Cu[N(CN)$_2$]Br}}

\begin{document}

\title{Superfluid density and microwave conductivity of FeSe superconductor:\\  ultra-long-lived quasiparticles and extended s-wave energy gap}

\author{Meng Li}
\affiliation{Department of Physics and Astronomy, University of British Columbia, Vancouver, BC, V6T 1Z1, Canada}
\author{N.~R.~Lee-Hone}
\affiliation{Department of Physics, Simon Fraser University, Burnaby, BC, V5A~1S6, Canada}
\author{Shun Chi}
\affiliation{Department of Physics and Astronomy, University of British Columbia, Vancouver, BC, V6T 1Z1, Canada}
\author{Ruixing Liang}
\author{W.~N.~Hardy}
\author{D.~A.~Bonn}
\affiliation{Department of Physics and Astronomy, University of British Columbia, Vancouver, BC, V6T 1Z1, Canada}
\affiliation{Canadian Institute for Advanced Research, Toronto, Ontario, MG5 1Z8, Canada}
\author{E.~Girt}
\affiliation{Department of Physics, Simon Fraser University, Burnaby, BC, V5A~1S6, Canada}
\author{D.~M.~Broun}
\affiliation{Department of Physics, Simon Fraser University, Burnaby, BC, V5A~1S6, Canada}
\affiliation{Canadian Institute for Advanced Research, Toronto, Ontario, MG5 1Z8, Canada}

\begin{abstract}
FeSe is an iron-based superconductor of immense current interest due to the large enhancements of \tc\ that occur when it is pressurized or grown as a single layer on an insulating substrate.  Here we report precision measurements of its superconducting electrodynamics, at frequencies of \mbox{202 and 658~MHz} and at temperatures down to 0.1~K.  The quasiparticle conductivity reveals a rapid collapse in scattering on entering the superconducting state that is strongly reminiscent of unconventional superconductors such as cuprates, organics and the heavy fermion material \cecoin.  At the lowest temperatures the quasiparticle mean free path exceeds 50~$\mu$m, a record for a compound superconductor.  From the superfluid response we confirm the importance of multiband superconductivity and reveal strong evidence for a finite energy-gap minimum.
\end{abstract}
  
\maketitle{} 

A recurring theme in correlated electron research is the sensitivity of such materials to small perturbations, which, when applied, can tune the material through a range of distinct electronic ground states.\cite{Sondhi:1997p360,Sachdev:2000p576,Lee:2000p993,Sachdev:2003p428,Coleman:2005p454,Broun:2008ci}  At the heart of this behaviour is a delicate balance between kinetic energy and potential energy, with kinetic energy favouring delocalized electrons and potential energy promoting various types of electronic order.   This balance can be tipped in one direction or the other by the application of magnetic field\cite{Custers:2003p472} and hydrostatic pressure,\cite{Mathur:1998ub} and by making small changes in chemical composition.\cite{vonLohneysen:1994cs,Yeh:2002eo}  In the iron-based superconductor FeSe, the first indication of this sensitivity occurs at around 90~K when the material changes from tetragonal to orthorhombic as it enters a nematic phase,\cite{McQueen:2009hs,Watson:2015kn} lowering rotational symmetry without breaking translational symmetry.  At lower temperatures FeSe becomes a superconductor, at $T_\mathrm{c} \approx 9$~K under ambient conditions,\cite{Hsu:2008ep} with \tc\ rapidly increasing by a factor of 4 under hydrostatic pressure.\cite{Medvedev:2009ex}  Even more dramatic enhancements occur in single-layer FeSe grown on insulating and semi-insulating substrates such as SrTiO$_3$, with reported values of \tc\ up to 100~K.\cite{QingYan:2012jx,He:2013cn,Ge:2014hc,Peng:2014dk}  A major effort is now underway to understand this fascinating new example of high temperature superconductivity.

FeSe is one of the simplest iron-based materials, superconducting at its stoichiometric composition and available as high quality single crystals grown using vapour transport methods.\cite{Bohmer:2013ee,Chareev:2013em}  A crucial question is to understand what the bulk, ambient superconductor is holding in reserve, that it can respond so strongly when pressurized\cite{IMAI:2009uu} or placed in contact with a substrate.\cite{Zhou:2016vk}  Central to addressing this issue is the identification of the superconducting gap structure and, ultimately, the pairing glue.\cite{IMAI:2009uu,Chubukov:2015jp,Zhang:2015wu,Mazin:2015im,Si:2016kl,Zhou:2016vk}  On the question of gap structure, a number of measurements emphasize the importance of multiband superconductivity,\cite{Khasanov:2010dx,Lin:2011fw,Lei:2012fu,Kasahara:2014gt,Jiao:2016td,Teknowijoyo:2016uf} but differ on whether or not the energy gap has nodes.  The main evidence for nodes appears in a combined study\cite{Kasahara:2014gt} that reports  large residual thermal conductivity and power-law penetration depth, $\lambda(T)$, although these conclusions are by no means universally agreed on.\cite{AbdelHafiez:2013eu,BourgeoisHope:2016ts,Teknowijoyo:2016uf}  A number of scanning tunnelling microscopy (STM) studies show evidence of V-shaped tunneling spectra,\cite{Song:2011em,Kasahara:2014gt} but the picture here is also complicated. Some STM studies report features that look more like finite gaps,\cite{Watashige:2015fa,Jiao:2016td} and there can be a complicated interplay with twin boundaries.\cite{Watashige:2015fa}  In addition, within experimental uncertainty, all STM studies appear to observe zero conductance at zero bias, something that can only occur if there is a single-particle gap at least three times the measurement temperature.  This on its own would imply a finite gap minimum in the range \mbox{$\Delta_\mathrm{min}/k_\mathrm{B} \approx 1.2$ to 4.5~K}.  Also weighing in in favour of a finite gap minimum are a number of bulk probes including heat capacity,\cite{Lin:2011fw,Jiao:2016td} thermal conductivity,\cite{BourgeoisHope:2016ts} $\mu$SR,\cite{Khasanov:2008ga} lower critical field\cite{AbdelHafiez:2013eu} and penetration depth.\cite{Teknowijoyo:2016uf}

In the work reported here we use precision low temperature measurements of the electrodynamic response of FeSe to shed further light on its gap structure and superconducting charge dynamics, in the process revealing extremely long-lived quasiparticles, a testament to the high degree of crystalline order that is possible in this  intriguing material.
\\

\begin{figure*}[ht]
\centering
\includegraphics*[width =\textwidth]{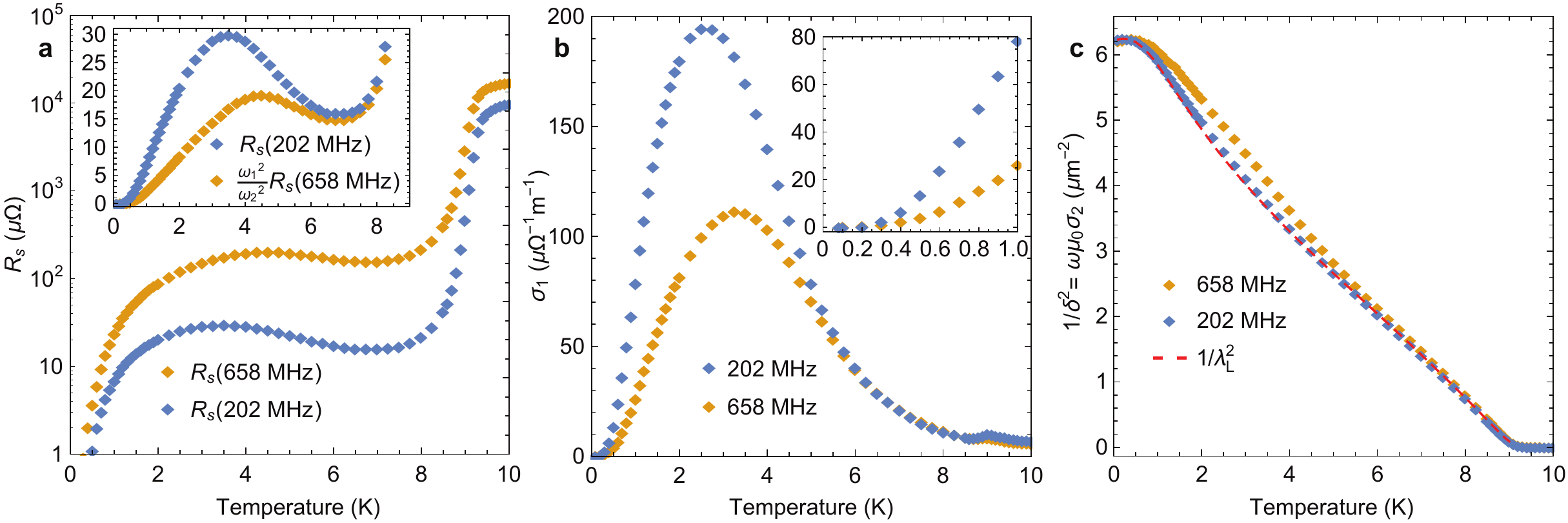}
\caption{{\bf Electrodynamic response of FeSe superconductor at 202 and 658 MHz.}   ({\bf a}) Surface resistance of FeSe with $R_\mathrm{s}$ on a logarithmic scale.  Inset: zoomed-in view of $R_\mathrm{s}$ in the superconducting state on a linear scale, with the expected $\omega^2$ frequency dependence of $R_\mathrm{s}$ scaled out at the higher frequency. ({\bf b}) Real part of the quasiparticle conductivity, $\sigma_1(T)$.  Inset: zoomed-in view showing the freeze-out of conductivity at low temperatures.  ({\bf c}) Frequency dependent superfluid fluid density, $1/\delta^2 \equiv \omega \mu_0 \sigma_2$.  The dashed line shows the London superfluid density, $1/\lambda_\mathrm{L}^2 \equiv 1/\delta^2(\omega \to 0)$. } 
\label{fig:Electrodynamics}
\end{figure*} 

\noindent \textsf{\textbf{Results}}

In the low frequency limit the electrodynamics of a superconductor are dominated by the superfluid response, giving rise to perfect dc conductivity.  Far from being a quiescent state, Cooper pairs are continually being broken apart into quasiparticle excitations then reforming in a phase-coherent manner --- such scattering does not degrade a steady current and is important in establishing the equilibrium superfluid density.\cite{Truncik:2013hr}  In order to study the electrodynamic response of the quasiparticles, higher frequencies are needed, ideally comparable to the quasiparticle relaxation rate.  Because the superfluid has finite inertia, a high frequency supercurrent is accompanied by an electric field, which in turn couples to the quasiparticle excitations and gives rise to a dissipative response.  This results in power absorption that increases as $\omega^2$ and is directly proportional to the real part of the quasiparticle conductivity, $\sigma_1$.\\

\noindent {\bf Surface impedance.}  The experimentally accessible quantity is the surface impedance, $Z_\mathrm{s} = R_\mathrm{s} + \I X_\mathrm{s}$, with $R_\mathrm{s} \approx \frac{1}{2} \omega^2 \mu_0^2 \lambda^3 \sigma_1$ and $X_\mathrm{s} \approx \omega \mu_0 \lambda$.  As we will show below the quasiparticle lifetimes in FeSe are extraordinarily long for a compound superconductor.  Experimentally, this means that the GHz-frequency resonators typically used in this type of measurement are too fast, not allowing the quasiparticles sufficient time to relax during the measurement period.  That said, surface resistance measurements face significant difficulties at lower frequencies and are rarely attempted:\cite{GRIMES:1991p824}  $R_\mathrm{s}$ falls off as $\omega^2$ and the characteristic size of resonators increases like $1/\omega$, making a mm-sized single crystal a negligible perturbation inside a large resonator volume.  Our solution has been to employ a special self-resonant coil, of diameter 4~mm and length 10~mm, wound from superconducting Nb wire and operating at $\omega_1/2 \pi = 202$~MHz and  $\omega_2/2 \pi = 658$~MHz, with empty-coil quality factors of several hundred thousand.  Such a resonator has sufficient sensitivity to resolve both the superfluid and quasiparticle response of a mm-sized crystal of FeSe.  

Figure~\ref{fig:Electrodynamics}a shows surface resistance data for FeSe at the two frequencies.  There is a sharp superconducting transition at $T_\mathrm{c} = 9.1$~K, indicative of a homogeneous sample.   Surface resistance drops quickly as Meissner screening currents take over from the normal-state skin effect, with $R_\mathrm{s}(T)$ reaching a minimum at $T \approx 7$~K.  Below this temperature the surface resistance rises again, with a peak that moves to lower temperature with decreasing frequency.  At the lowest temperatures $R_\mathrm{s}$ falls again, decreasing into the $\mu\Omega$ range.  The nonmonotonic temperature dependence of $R_\mathrm{s}$ has been observed in only one other material system --- ultrahigh purity \ybco{6+x} --- and is immediately indicative of a system with rapidly collapsing scattering and extremely long-lived quasiparticles.\cite{Bonn:1992fx,Kamal:1998et,Hosseini:1999p383,Harris:2006p388}  The main  differences between FeSe and \ybco{6+x} are twofold: the details of the freeze-out of $R_\mathrm{s}(T)$ at low temperatures suggest a finite energy gap in FeSe, instead of the symmetry-protected \dwave\ nodes in \ybco{6+x};\cite{HARDY:1993p632,Bonn:2006ue} and in FeSe the nonmonotonic $R_\mathrm{s}(T)$ appears only below about 1~GHz, as compared to about 70~GHz in \ybco{6.993}.\cite{Hosseini:1999p383,Harris:2006p388}  This on its own suggests that quasiparticles lifetimes are unusually long in FeSe.\\

\noindent {\bf Microwave conductivity.}  From the surface impedance we obtain the high frequency conductivity $\sigma = \sigma_1 - \I \sigma_2$.  The real part of the conductivity is plotted in Fig.~\ref{fig:Electrodynamics}b for the two measurement frequencies and shows more clearly the underlying quasiparticle dynamics that are responsible for the nonmonotonic form of $R_\mathrm{s}(T)$.  On cooling through \tc, $\sigma_1(T)$ starts to rise.  In this temperature range $\sigma_1$ shows no appreciable frequency dependence, consistent with a quasiparticle relaxation rate that is much larger than the measurement frequencies.  On further decreasing temperature $\sigma_1(T)$ rises rapidly, eventually peaking at around a third of \tc, at a value 40 times higher than $\sigma_1(T_\mathrm{c})$ in the case of the 202~MHz data.  As well as in the \ybco{6+x} system\cite{Nuss:1991wj,Bonn:1992fx,Hosseini:1999p383, Harris:2006p388} qualitatively similar behaviour is observed in the heavy fermion superconductor \cecoin,\cite{Ormeno:2002p404, Truncik:2013hr} and the organic superconductor \Br,\cite{Milbradt:2013ic} although only in Ortho-I \ybco{6.993} is the enhancement as dramatic.  In all cases the physics is the same --- on passing through \tc\ there is a sudden collapse in quasiparticle scattering that vastly outpaces the steadier condensation of quasiparticles into the superfluid condensate.  Consistent with this conclusion $\sigma_1$  develops substantial frequency dependence below $T_\mathrm{c}/2$, indicating that the quasiparticle relaxation rate is becoming comparable to the measurement frequency --- in the case of FeSe, a relaxation rate in the sub-GHz range.  
\\

\noindent {\bf Superfluid density.}  The imaginary part of the conductivity is dominated by the superfluid term, $\sigma_\mathrm{s} = 1/\I \omega \mu_0 \lambda_\mathrm{L}^2$, where $\lambda_\mathrm{L}$ is the London penetration depth.  We plot the frequency-dependent superfluid density, $1/\delta(\omega)^2 \equiv \omega \mu_0 \sigma_2(\omega)$, in Fig.~\ref{fig:Electrodynamics}c.  There is some frequency dependence in the lower part of the temperature range, consistent with the presence of slowly relaxing quasiparticles below $T_\mathrm{c}/2$ that eventually freeze out at the lowest temperatures.  Fits to a generalized two-fluid model (see Methods) are used to extrapolate $\omega \mu_0 \sigma_2(\omega)$ to the static limit and obtain the London superfluid density, $1/\lambda_\mathrm{L}^2$.  The superfluid density shows a strong, approximately linear temperature dependence over most of the temperature range, with upwards curvature in  $1/\lambda_\mathrm{L}^2(T)$ at around $T_\mathrm{c}/3$, which we will discuss below in the context of multiband superconductivity.  At the lowest temperatures there is a substantial flattening of $1/\lambda_\mathrm{L}^2(T)$, similar to that seen in $R_\mathrm{s}(T)$ and $\sigma_1(T)$, that we will show is evidence of small but finite gap minima in one of the bands.  Broadly similar behaviour in $1/\lambda_\mathrm{L}^2(T)$ has very recently been reported in Ref.~\onlinecite{Teknowijoyo:2016uf}.
\\

\noindent \textsf{\textbf{Discussion}}\\

\begin{figure}[t]
\centering
\includegraphics[width = 0.8 \columnwidth]{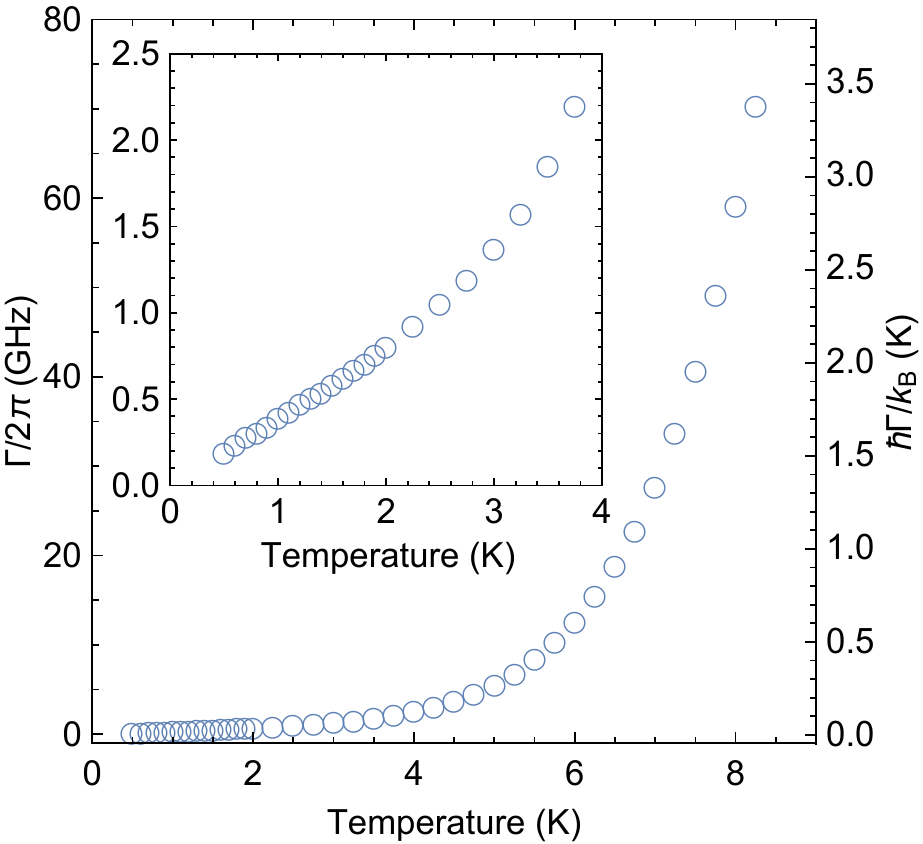}
\caption{{\bf Quasiparticle relaxation dynamics.}  The quasiparticle relaxation rate, $\Gamma(T)$, is obtained from the fitting to conductivity with a modified Drude model.  For reference, the normal-state relaxation rate is approximately 135~GHz at 12~K.   Inset: low temperature zoom, revealing an approximately linear temperature dependence of $\Gamma$ between \mbox{$T = 0.5$ and 2.5~K}.} 
\label{fig:Gamma}
\end{figure}

To gain further insight into the superconducting charge dynamics we have fit a generalized two-fluid model (see Methods) to the complex conductivity: at each temperature this gives the average quasiparticle relaxation rate, $\Gamma$, and the London superfluid density.  The relaxation rate is plotted in Fig.~\ref{fig:Gamma}.  As expected from the qualitative behaviour of  $R_\mathrm{s}(\omega,T)$ and $\sigma_1(\omega,T)$ there is a rapid collapse in $\Gamma(T)$ on cooling into the superconducting state, which in the context of the cuprates, organics and \cecoin\ is interpreted as a strong indication that the fluctuations responsible for inelastic scattering are electronic in origin and are gapped by the onset of superconductivity.  Another factor that is relevant to the suppression of $\Gamma(T)$ in these systems is the importance of Umklapp processes in relaxing electrical currents\cite{Walker:2000ux,Duffy:2001dg} --- as we will discuss later, energy conservation in Umklapp events becomes difficult to satisfy in multiband superconductors with anisotropic gap.  Below 2.5~K $\Gamma(T)$ locks into a linear temperature dependence.  Such behaviour is reminiscent of cuprate superconductors in the Born-scattering limit,\cite{HIRSCHFELD:1994p570,Schachinger:2003p635,Turner:2003p331,Ozcan:2006bm} where the linear temperature dependence of $\Gamma$ reflects the linear-in-energy density of states (DOS) of the \mbox{\dwave}\ quasiparticles.  
\begin{figure*}[htb]
\centering
\includegraphics*[width = 0.75 \textwidth]{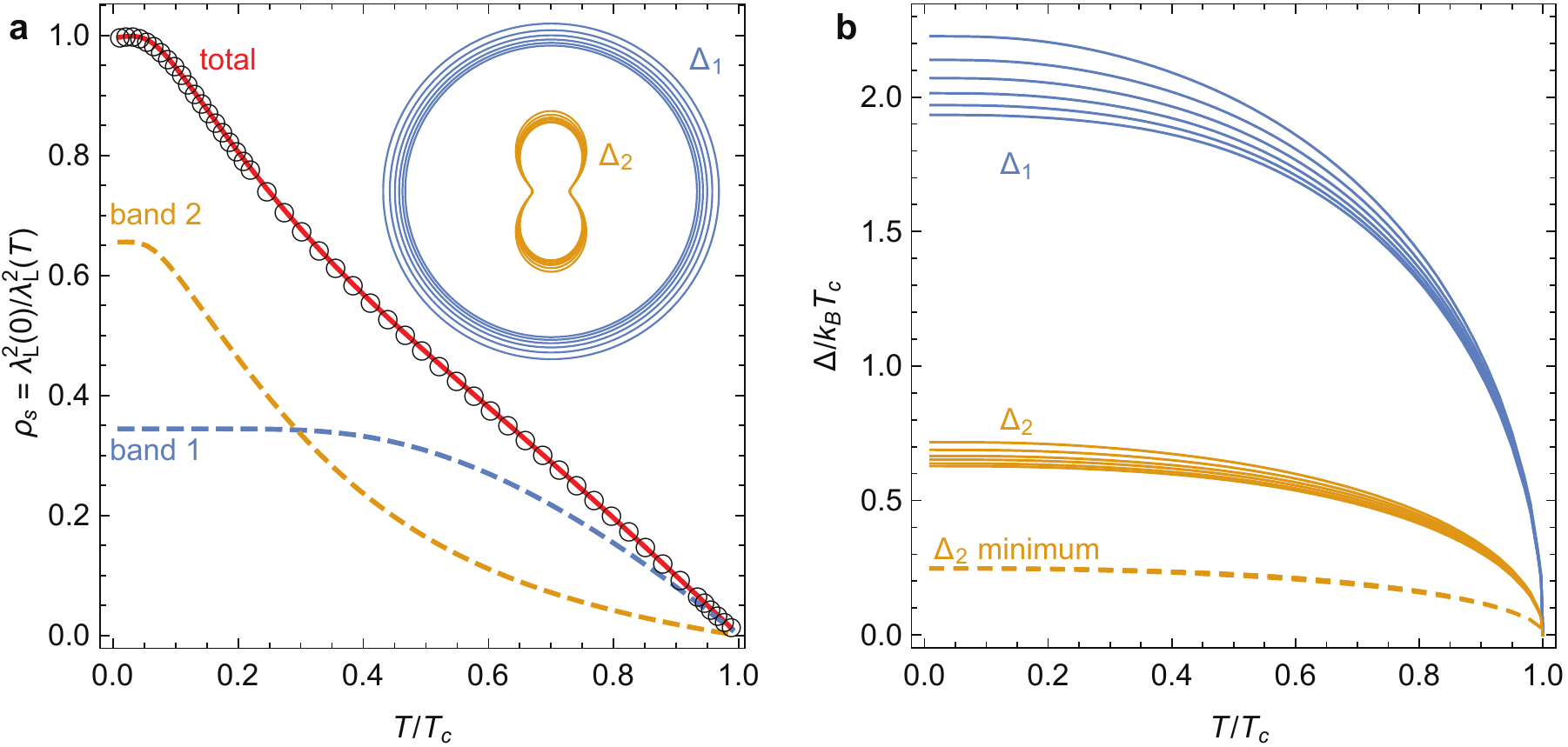}
\caption{{\bf Superfluid density in a two-band extended s-wave model.}   ({\bf a}) Superfluid density is calculated using a two-band extended s-wave model and fit to the London superfluid density, $1/\lambda_\mathrm{L}^2(T)$.  Inset: A polar plot showing the schematic form for the two-band extended s-wave gap at zero temperature, for various values of the DOS parameter, \mbox{$0.25 < n_1 < 0.5$}.  ({\bf b}) Temperature dependence of the rms gap amplitudes on bands one and two, and the overall gap minimum, for the same range of $n_1$.} 
\label{fig:TwoBand}
\end{figure*} 
We will argue below that the energy gap in FeSe has an extended \swave\ form, with deep gap-minima --- such a state is  accompanied by a linear DOS at energies immediately above the gap minimum and possibly leads to $\Gamma(T) \propto T$ in the intermediate temperature range.  Below $T = 0.5$~K, there is insufficient quasiparticle conductivity to carry out accurate fits to the two-fluid model, so $\Gamma$ is not plotted in this range.  At the lowest temperatures the scattering rate is $\Gamma_\mathrm{min}/2 \pi \approx 200$~MHz, corresponding to $\hbar \Gamma_\mathrm{min}/k_\mathrm{B} \approx 10$~mK in temperature units.  Using a value of $v_\mathrm{F} \approx 7 \times 10^4$~m/s for the Fermi velocity,\cite{Watson:2015kn}  the quasiparticle mean free path is $\ell_0 = v_F/\Gamma_\mathrm{min} \approx 55~\mu$m.  This is the largest value we are aware of for a compound superconductor and is 5 times larger than that of the best \ybco{6.52} crystals, where the low temperature scattering rate reaches $\Gamma/2\pi \approx 3.3$~GHz.\cite{Turner:2003p331}  On the experimental side, we emphasize that this result is on a firm footing, as our lowest measurement frequency is of the same order as $\Gamma_\mathrm{min}/2 \pi$.  Although $\ell_0$ is larger than almost any other length scale in our sample, nonlocal effects are likely suppressed by four different considerations: the larger-than-normal in-plane penetration depth in FeSe\cite{Khasanov:2008ga,AbdelHafiez:2013eu,Kasahara:2014gt} ($\lambda_0 \approx 400$ to 500~nm); the quasi-2D electronic structure,\cite{Subedi:2008hc,Terashima:2014ft,Watson:2015kn,Mukherjee:2015kb} which, in our experimental geometry, forces quasiparticles to propagate at low angles to the sample surface; the vanishing of quasiparticle group velocity near the gap minima; and the diffusive nature of small-angle scattering in a superconductor with anisotropic gap.\cite{Durst:2000p963}  The last two points work together to make the distance travelled by the quasiparticle wavepacket between large-angle scattering events much smaller than the apparent mean free path.  Even taking into account the significant enhancements of electrical mean free path due to these effects,  the inferred value of $\ell_0$ is remarkable and suggests that FeSe may provide a novel testing ground for exploring hydrodynamic effects\cite{Zaanen:2016hn,Moll:2016ju} in superconducting transport.

We turn now to the superfluid density, a thermodynamic probe sensitive to the itinerant electronic degrees of freedom, the temperature dependence of which is controlled by the energy dependence of the quasiparticle DOS.\cite{Prozorov:2006wr}  The main features of $1/\lambda_\mathrm{L}^2$, pointed out above, are its strong temperature dependence across most of the superconducting range, indicating a broad distribution of energy scales in the gap; the upwards curvature of $1/\lambda_\mathrm{L}^2(T)$ around $T_\mathrm{c}/3$, a hallmark of multiband physics;\cite{Kogan:2009ew} and its pronounced flattening at low temperatures, indicating the presence of finite gap minima instead of true nodes.\cite{Prozorov:2006wr}  To put these observations on a quantitative footing we have fit the normalized superfluid density, $\rho_\mathrm{s} = \lambda_\mathrm{L}^2(0)/\lambda_\mathrm{L}^2(T)$, to a two-band model similar to that developed by Kogan, Martin and Prozorov (KMP) in Ref.~\onlinecite{Kogan:2009ew}.  From initial fitting attempts it became apparent that the small energy scale implied by our data could not be captured by a two-band model with isotropic gaps, motivating us to modify the KMP formalism to include gap anisotropy of an extended \swave\ form with $\Delta(\phi) \propto \left(1 + \sqrt{2} \alpha \cos(2 \phi)\right)$, as summarized in Methods.  In our phenomenological model the gap anisotropy is put in by hand but naturally arises from orbital-dependent effects in microscopic models.\cite{Mukherjee:2015kb}  Our fitting parameters are the DOS imbalance between the two bands (parameterized by $n_1 = N_1/N_\mathrm{total}$); the coupling constants $\lambda_{11}, \lambda_{22}$ and $\lambda_{12} = \lambda_{21}$, which represent intraband and interband pairing, respectively; the gap anisotropy parameters, $\alpha_1$ and $\alpha_2$; and a factor $\gamma$ that controls the relative superfluid weights.   Early on in this process we observed that the large-gap band seemed to require little anisotropy, so $\alpha_1$ was set to zero in the remaining work.  Equally good fits to $\rho_\mathrm{s}(T)$ are obtained in the range \mbox{$0.25 < n_1 < 0.5$}, with the fit for $n_1 = 0.35$ shown in Fig.~\ref{fig:TwoBand}a.  A typical value of the anisotropy parameter for band 2 is $\alpha_2 = 0.42$.  The schematic forms of the energy gaps are shown as functions of angle and temperature in Fig.~\ref{fig:TwoBand} for \mbox{$0.25 < n_1 < 0.5$}.  Note that while there is some variation of $\Delta_1$ and $\Delta_2$ with the choice of $n_1$, all fits agree well on the minimum energy gap, $\Delta_\mathrm{min}/k_\mathrm{B} \approx 0.25~T_\mathrm{c} = 2.3$~K, which is a factor of 8 smaller than the large gap in band~1.  (Interestingly, very similar conclusions have independently been reached by a field-dependent thermal conductivity study carried out on similar crystals, \cite{BourgeoisHope:2016ts} and from penetration-depth experiments,\cite{Teknowijoyo:2016uf} where fits to a one-band extended \swave\ model estimate $\Delta_\mathrm{min}/k_\mathrm{B} \approx 0.3~T_\mathrm{c}$.)
Within our multiband interpretation, the large conductivity peaks seen Fig.~\ref{fig:Electrodynamics}b are the result of long-lived quasiparticles that are thermally excited in the vicinity of the deep gap minima in band~2.  This has two important consequences for the relaxation dynamics:  there will be a strong reduction in the phase space for recoil when low energy quasiparticles undergo elastic impurity scattering;\cite{HIRSCHFELD:1994p570,Durst:2000p963,Schachinger:2003p635} and Umklapp events,\cite{Walker:2000ux,Duffy:2001dg} which are necessary if inelastic processes are to relax the electrical current, will require the low energy quasiparticles to partner with quasiparticles on the large-gap band, and these excitations will be strongly gapped.  

We emphasize that while the two-band model is used mainly for illustrative purposes, the importance of multiband effects and the presence of finite gap minima --- also visible in $R_\mathrm{s}(T)$ and $\sigma_1(T)$ --- are robust conclusions.  In fact, each of the gap values can be directly tied to qualitative features in the temperature dependence of the superfluid density: the gap minimum, $\Delta_\mathrm{min}$, is fixed by the small range of temperatures over which thermally activated behaviour is observed; $\Delta_2$, the average value of the gap in the subdominant band, is linked to the upwards curvature in $\rho_\mathrm{s}(T)$, which occurs in a range near $T_\mathrm{c}/3$; and the dominant gap, $\Delta_1$, is set by \tc\ itself.  It is  worth mentioning that the factor of 3 ratio between $\Delta_1$ and $\Delta_2$ is difficult to obtain in purely repulsive models of pairing, as the dominant interaction in that case is the interband pairing,\cite{Kogan:2009ew} and an unrealistically large imbalance of the density of states ($n_1 \approx 0.05$) must then be assumed to obtain $\Delta_1 \approx 3 \Delta_2$.  

The observation of activated exponential temperature dependence in our low temperature data and the concomitant identification of finite gap minima are at odds with reports of line nodes in FeSe, in particular the measurements of superfluid density and thermal conductivity in Ref.~\onlinecite{Kasahara:2014gt}. As pointed out in the introduction, several tunnelling spectroscopy experiments present V-shaped spectra that at first sight appear to be indicative of gap nodes,\cite{Song:2011em,Kasahara:2014gt} although the finite-temperature effects (in particular, nonzero conductance at zero bias) that would be expected in the nodal case are not reported.  Instead, our observations are in close keeping with the conclusions of a recent thermal conductivity study carried out on the same samples as ours,\cite{BourgeoisHope:2016ts}  and with heat capacity,\cite{Lin:2011fw} lower critical field\cite{AbdelHafiez:2013eu} and penetration depth studies.\cite{Teknowijoyo:2016uf}  While it is tempting to attribute the lifting of accidental gap nodes to impurity scattering,\cite{BourgeoisHope:2016ts,Teknowijoyo:2016uf} which should homogenize the energy gap in an anisotropic \swave\ superconductor, we find this hard to reconcile with the exceptionally long quasiparticle mean free path reported here.  Nevertheless, sample-to-sample variations of physical properties usually have a microstructural origin.  In the case of FeSe,  a  possible alternative to point-like disorder is the presence of twin boundaries, which form in the nematic phase below 90~K.\cite{McQueen:2009hs,Watson:2015kn,Tanatar:2015tu,Teknowijoyo:2016uf}  These have been shown to have a significant effect on the spatial variation of the energy gap in a recent scanning tunneling spectroscopy experiment.\cite{Watashige:2015fa}  Further insights into this may come from local-probe experiments carried out at lower temperatures, if the tunnel junction can be sufficiently cooled, and from measurements on single-domain, macroscopically detwinned samples, when these become available.
 \\

\noindent \textsf{\textbf{Methods}}\\
\footnotesize

\noindent \textsf{\textbf{Samples.}}  High quality single crystals of FeSe were grown by vapour transport using the method described in Ref.~\onlinecite{Bohmer:2013ee}.  The sample used for the microwave conductivity experiment was a mm-sized platelet, cleaved from a thicker crystal to have a thickness along the $c$~direction of $t = 15~\mu$m.\\ 

\noindent \textsf{\textbf{Surface impedance and microwave conductivity.}} Measurements of surface impedance, $Z_\mathrm{s} = R_\mathrm{s} + \I X_\mathrm{s}$, were made at two frequencies using cavity perturbation\cite{1946Natur.158..234P,Altshuler:1963wq,Klein:1993p1129,Donovan:1993p1114,Huttema:2006p344,Bonn:2007hl} of a self-resonant coil wound from Nb wire and housed inside a Pb:Sn-coated enclosure mounted below the mixing chamber of an MX40 $^3$He--$^4$He dilution refrigerator.\cite{Truncik:2013hr}  During the experiments, the resonator was maintained at a fixed temperature of 1.5~K, while the sample temperature was scanned between $T= 0.1$ and 20~K using a silicon hot-finger\cite{Sridhar:1988p495} thermally linked to the mixing chamber.  Data were obtained at 202~MHz using the fundamental mode of the coil resonator,  and at 658~MHz using its second overtone.  In both cases, the sample was positioned at a local maximum of the microwave magnetic field, $H_\mathrm{rf}$, which was applied parallel to the FeSe layers to induce predominantly in-plane screening currents.  Measurements were carried out under conditions of constant $H_\mathrm{rf}$, to avoid nonlinearities, and microwave power was kept low enough to prevent self-heating.  Temperature-dependent changes in the effective surface impedance, $Z_\mathrm{s}^\mathrm{eff}$,  are obtained directly from shifts in resonator frequency, $f_0$, and resonant bandwidth, $f_\mathrm{B}$, using the cavity perturbation approximation \mbox{$\Delta Z_\mathrm{s}^\mathrm{eff} = \Gamma \left( \Delta f_\mathrm{B}(T)/2 - \I \Delta f_0(T)\right)$}.  Here $\Gamma$ is a resonator constant determined empirically from the DC resistivity of FeSe samples from the same batch,\cite{BourgeoisHope:2016ts} and $\Delta f_\mathrm{B}$ is the change in resonator bandwidth with and without the sample.  Due to the thin sample and  low measurement frequencies, finite-size effects are important near and above \tc\ and are corrected using the 1D finite-size formula $Z_\mathrm{s}^\mathrm{eff} = Z_\mathrm{s} \tanh(\I \omega \mu_0 t/2 Z_\mathrm{s})$, where $Z_\mathrm{s} $ is the surface impedance of a semi-infinite sample.  The absolute zero-temperature surface reactance, $X_\mathrm{s}(0) = \omega \mu_0 \lambda_0$, is set using a previously reported value of the zero-temperature penetration depth, $\lambda_0 = 400$~nm.\cite{Kasahara:2014gt} The complex microwave conductivity, $\sigma = \sigma_1 - \I \sigma_2$, is obtained from the surface impedance using the local electrodynamic relation, $\sigma = \I \omega \mu_0/Z_\mathrm{s}^2$.\\
 
\noindent \textsf{\textbf{Generalized two-fluid model and modified Drude conductivity.}} For the purposes of extracting relaxation rate $\Gamma$ and extrapolating superfluid density to the zero-frequency limit we fit conductivity data to a generalized two-fluid model\cite{Ozcan:2006bm}
\begin{equation}
\sigma = \frac{ne^2}{m^\ast} \left[\frac{f_\mathrm{s}}{\I \omega} + \frac{f_\mathrm{n}}{\Gamma} \left(\frac{1}{1 + (\omega/\Gamma^\prime)^y} - \I\;\mathrm{KK}(\omega/\Gamma^\prime,y)\right)\right]\;,
\end{equation}
where $\Gamma^\prime = \Gamma \times \frac{y}{2} \sin(\frac{\pi}{y})$ is a convenient scaling that makes the integrated quasiparticle spectral weight independent of $y$.   The imaginary part of the quasiparticle conductivity is obtained using a Kramers--Kr\"onig transform\cite{Ozcan:2006bm} and $f_\mathrm{s} + f_\mathrm{n} = 1$ in order to conserve spectral weight. Away from the Drude limit $(y =2)$ the frequency exponent $y$ modifies the quasiparticle conductivity to better capture situations with strongly energy-dependent scattering.\cite{Turner:2003p331,Truncik:2013hr}   In FeSe at low temperatures the best-fit value was found to be $y=1.5$; at high temperatures, where $\sigma_1$ has little frequency dependence, the fits are insensitive to the choice of $y$: the frequency exponent was subsequently fixed at  $y=1.5$ at all temperatures.  The static, London superfluid density is $1/\lambda_\mathrm{L}^2 = f_\mathrm{s} \mu_0 n e^2/m^\ast$.\\

\noindent \textsf{\textbf{Two-band extended s-wave superconductor.}} To model the superfluid density,  calculations are made for a clean-limit, two-band superconductor in which the angle dependence of the superconducting order parameter in bands $\nu = 1,2$ has the form $\Delta_\nu(T) \Omega_\nu(\phi)$, where $\Omega_\nu(\phi) = \left(1 + \sqrt{2} \alpha_\nu \cos(2 \phi)\right)/\sqrt{1 + \alpha_\nu^2}$ takes an extended s-wave form, normalized so that $\langle \Omega_\nu^2(\phi)\rangle_\phi = 1$.   Here $\alpha_\nu$ controls the degree of gap anisotropy, with gap nodes appearing for $\alpha_\nu \ge 1/\sqrt{2}$. We follow the weak-coupling Eilenberger method described in Ref.~\onlinecite{Kogan:2009ew}, modified to incorporate angle-dependent order parameters.  Specifically, we solve the self-consistent gap equation
\begin{equation}
\Delta_\nu = \sum_{\mu = 1,2} \!\!n_\mu \lambda_{\nu\mu} \Delta_\mu\!\!\sum_{\omega_n>0}^{\omega_D}  \left\langle\frac{ 2 \pi k_\mathrm{B} T\; \Omega_\mu^2(\phi)}{\sqrt{\Delta_\mu^2 \Omega_\mu^2(\phi)\!+\!\hbar^2 \omega_n^2}} \right\rangle_{\!\!\phi},
\label{Eq:TwoBandGapEquationReduced}
\end{equation}
where the $\omega_n$ are the fermionic Matsubara frequencies and the frequency cut-off $\omega_D$ is set by the measured \tc.  The relative densities of states  $n_\nu$ and the coupling constants $\lambda_{\nu\mu}$ have the same form as in Ref.~\onlinecite{Kogan:2009ew} and, subject to the constraints $n_1 + n_2 = 1$ and $\lambda_{\nu\mu} = \lambda_{\mu\nu}$, are adjustable parameters. The normalized superfluid density, $\rho_\mathrm{s} = \lambda^2(0)/\lambda^2(T)$, is directly computed from $\Delta_1(T)$ and $\Delta_2(T)$ and is a simple linear combination of the contribution from each band: $\rho_\mathrm{s}(T) = \gamma \rho_1(T) + (1 - \gamma) \rho_2(T)$, where the parameter $\gamma$ controls the relative magnitude of the superfluid weights.

\normalsize


\vspace{3 mm}
\normalsize
\noindent \textsf{\textbf{Acknowlegdements}}\\
\footnotesize
We thank A.~V.~Chubukov, J.~S.~Dodge, N.~Doiron-Leyraud, A.~V.~Frolov, P.~J.~Hirschfeld, M.~P.~Kennett	 and L.~Taillefer for discussions and correspondence. Research support was provided by the Natural Science and Engineering Research Council of Canada, the Canadian Institute for Advanced Research, and the Canadian Foundation for Innovation.\\

\normalsize
\noindent \textsf{\textbf{Author contributions}}\\
\footnotesize
M.L.\ and D.A.B.\ carried out the initial measurements of surface resistance and penetration depth. N.L.H.\ and D.M.B.\ designed and set up the dilution-refrigerator-based coil-resonator system and carried out the low temperature surface impedance measurements.   S.C., R.L, W.N.H.\ and D.A.B\ contributed to the growth of FeSe crystals.  D.M.B.\ and N.L.H.\ carried out numerical work for the two-band modelling.  D.A.B.\ and D.M.B.\ wrote the paper.  D.A.B., E.G.\ and D.M.B.\ supervised the project.\\ 

\normalsize
\noindent \textsf{\textbf{Additional information}}\\
\footnotesize
{\bf Competing financial interests.} The authors declare no competing financial interests.\\

\normalsize
\noindent \textsf{\textbf{Corresponding author}}\\
\footnotesize
Correspondence and requests for materials should be addressed to D.M.B. (email: dbroun@sfu.ca).\\

\normalsize

\end{document}